    Reply to the Report of the First Referee

                          List of the change made.

Remarks concerning applicability of the second-order perturbation theory
approximation are added just after Eq. 2 of the new version of the paper. 

\documentstyle[aps]{revtex}
 
\begin{document}
 \draft
 \title{Zero-bias anomalies of point contact resistance due to
 adiabatic electron renormalization of dynamical defects}
 \author{V.I.Kozub$^{1,2}$ and A.M.Rudin$^{1,3}
\footnote{Present address: Theoretical Physics Institute, University of
Minnesota, Minnneapolis, MN 55455}$}
 \address{
 $^{1}$A.F.Ioffe Physico-Technical Institute, St.-Petersburg, Russia \\
$^{2}$ Delft Institute of Microelectronics and Submicron Technology, TU
 Delft, 2628 CJ Delft, The Netherlands\\
$^{3}$ Institut f\"ur Festk\"orperforschung, Forschungszentrum
 J\"ulich, D52425 J\"ulich, Germany}
  \maketitle
 \begin{abstract}
We study effect of the  adiabatic electron renormalization on the parameters 
of the dynamical defects in the ballistic
metallic point contact. The upper energy states of the
``dressed'' defect are shown to give a smaller contribution to a
resistance of the contact than the lower energy ones. This holds both for the 
"classical" renormalization related to defect coupling with
average local electron density and for the "mesoscopic" renormalization
 caused by the   mesoscopic fluctuations of electronic
density the dynamical defects are coupled with.  In the case of mesoscopic
renormalization  one may treat the 
dynamical defect as coupled  with Friedel oscillations originated by the 
other  defects, both static and mobile.   Such coupling  lifts  the energy 
degeneracy of
the  states of the dynamical defects giving different mesoscopic contribution 
to resistance, 
and provides a new model for the
fluctuator  as for the object originated by  the electronic mesoscopic 
disorder rather than by the structural one.  The correlation between the 
defect energy and the defect contribution to the resistance  leads to 
zero-temperature and
zero-bias anomalies of the point contact resistance. 

A comparison of these anomalies with those predicted by the Two Channel Kondo
Model (TCKM) is made. It is shown, that although the proposed model is based on
a completely different from TCKM physical background, it leads to a zero-bias
anomalies of the point contact resistance, which are  qualitatively similar to
TCKM predictions.
 \end{abstract}
 
 \narrowtext
 \section{Introduction}
 Recent advances in nanofabrication technology have made it possible
 to visualize single defects with internal degrees of freedom -
 ``fluctuators'' \cite{Rogers84,Ralls88}, which lead to a
 ``telegraph'' resistance noise of nanometer scale systems. In metals
 these defects are believed\cite{Kogan84,K84a,K84b,K84c,Galp89} to be
 some structural defects,
 which at low temperatures are seen as the well-known two-level
 tunneling states (TLS) \cite{AHVP}.
 TLSs are the objects typical for strongly disordered amorphous solids
 \cite{AHVP} that switch by tunneling between their two possible
 configurations. 
 Although the microscopic nature of the fluctuators stays unclear
 (especially for ballistic devices made of pure metals), the
 experiments allow to study various phenomenological
 parameters of these objects, in particular, the interlevel
 spacing. 
 
 Kondo \cite{Kondo} was the first, who has pointed out that in metals
the parameters of the dynamical defects are strongly renormalized by
electrons.
 One can discriminate between adiabatic electron ``dressing'', which
 is related
 to a static electron response on the defect potential, and
 non-adiabatic one, which affects the tunneling process and leads to a
 renormalization of TLS tunneling matrix element (``dissipative
 tunneling''). It is this non-adiabatic effect that has attracted most
 attention (see e.g. Ref. \cite{Dissreview}) due to its evident
 importance for the defect dynamics. As for the adiabatic renormalization
 of the defect parameters, it is customary to include it into the
 bare values. This procedure is usually justified by the fact that
 the adiabatic effects are related to a responce of the whole electron
 systems while only a small strip of electron energies, close to the Fermi
 surface, is responcible for  transport properties and  sensitive
 to external factors like temperature or fields applied.
 However, as it was first demonstrated in Refs.\cite{K84a,K84c}, the
 adiabatic ``dressing'' of the fluctuator, in particular, the
 adiabatic renormalization of the fluctuator energy splitting $E$, can be
 important and depends on the state of the electron system (e.g. on
 superconducting properties). Furthermore, very recently the surprizing
"magnetic tuning" of the TLS interlevel spacing observed\cite{Zimm}  for
TLSs in Bi nanoconstrictions was explained\cite{nondiss} as a direct result
of the adiabatic renormalization of TLS parameters by electrons, which
states are affected strongly by the magnetic field. 

 The purpose of the present paper is to
 show that {\it the adiabatically renormalized energy of the fluctuator
state
 correlates with the fluctuator contribution to the resistance}.
 Namely, the conductance is {\it larger} for the
 fluctuator on its {\it upper} level.  This fact causes, in
 particular, Kondo-like zero-temperature and zero-bias anomalies in
 differential conductance of the metallic point contacts.  Indeed, an
 increase of temperature leads, in average, to an increase of
 occupation numbers of the fluctuator upper levels. The abovementioned
 correlation causes a corresponding conductance {\it increase with
 temperature}, which imitates the Kondo-like behavior.  The same holds
 for an applied bias increase.
 
\section{Adiabatic renormalization of the parameters of the dynamical
defects in metals}

 Two mechanisms of the adiabatic renormalization can be considered.
 First one is due to possible difference of electron-fluctuator
 coupling potentials $V$ for the two of the fluctuator states
 ($|V^{(1)}| \neq |V^{(2)}|$) and was  studied in
 Ref.\cite{K84a,K84c,nondiss}.
 We will show that for this mechanism (which will be referred to as
 ``classical'' one) the abovementioned correlation is due to the fact
 that expressions for  the conductance and for the electron contribution to
defect
 energy include the same strength of the electron-defect coupling.
 
 Second mechanism of adiabatic renormalization of dynamical
 defects parameters was
 suggested by Altshuler and Spivak \cite{Alt89}. It implies mesoscopic
 electron density fluctuations, which lead to a difference, even for
 $|V^{(1)}| = |V^{(2)}|$ , of electron-fluctuator coupling strengths
 for different fluctuator states due to their spatial resolution. The
 same correlation  does occur for this ``mesoscopic''
 contribution, as for the ``classical'' one. We will show it for the
 experimentally important situation of ballistic point contact, where
 a description in terms of ``local'' interference, which involves
 finite number of scatterers, is possible.
 
 \subsection{Renormalization due to difference in defect-electron
 coupling potentials}
 
 Let us start from the ``classical'' effect. TLS energy splitting $E$
depends\cite{Galp89} on the TLS tunneling parameter $\Delta_0$ and on the TLS
asymmetry $\Delta$: 
 \begin{equation}
E= \sqrt{\Delta_0^2+\Delta^2}.
 \end{equation}
Interaction of TLS with conduction electrons leads to the renormalization of
both TLS asymmetry  \cite{K84a,K84c,nondiss} and of tunneling parameter
\cite{Kondo}. For the asymmetrical TLSs with the large enough barrier,
$\Delta \gg \Delta_0$, the renormalization of TLS asymmetry gives the
major contribution to $E$. As a result
 \cite{K84a,K84c,nondiss}, the electron-TLS coupling leads to the
renormalization of
 TLS ``bare'' energy splitting $E$ which has  a 
 form $E \rightarrow E + E_{el} $, where $E_{el} = E_{el,2} -
 E_{el,1}$,
 \begin{equation}\label{E_eli}
 E_{el,i} = \sum_{\bf k, {\bf k}'} \mid V_0^{(i)} \mid ^2
 {f(\varepsilon_{\bf k}) - f(\varepsilon_{{\bf k}'})
 \over  \varepsilon_{\bf k} - \varepsilon_{{\bf k}'}}  \approx
 - {|V_0^{(i)}|^2 \over \varepsilon_F } \; .
 \end{equation}
 Here $V_0^{(i)}$ is the electron-TLS coupling constant for $i$-th
 TLS configuration. The electron bandwidth is assumed to be of order
 of $\varepsilon_F$, and the electron distribution
 $f(\varepsilon)$ to have an equilibrium Fermi form $f(\varepsilon) =
 f_0 (\varepsilon ) = \{\exp [(\varepsilon - \varepsilon_F)/T] + 1
 \}^{-1}$. Applicability of the second-order perturbation theory approximation
is justified for Eq. (\ref{E_eli}) when $V_0^{(i)} \ll \varepsilon_F$. In
contrast to Kondo-like corrections,  Eq. (\ref{E_eli})  depends weakly on
temperature. The
renormalization $E_{el}$ is
 due to the difference in the values of the total electron energy,
 renormalized by a presence of TLS, for the TLS in states 1 and 2,
 respectively.  It is important, that $E_{el}$ is not, due to
 adiabaticity, sensitive to the details of inter-state transition
 mechanism.  Thus the problem is reduced to estimates of energies
 $E_{el,i}$ corresponding to different configurations of the defect
 which for this case can be considered as its independent
 realizations. Therefore, Eq. (\ref{E_eli}) holds both at low
 temperatures, when the fluctuator transitions are due to tunneling,
 and at higher temperatures when thermal activation dominates
 \cite{Galp89,KR93}.
 
 According to Eq.(\ref{E_eli}) the energy $E$ is {\it lowered} from
 the ``bare'' value. The lowering is larger the stronger is the
 coupling of the state with the electrons.  On the other hand, a
 presence of the defect in the state $i$ inside a ballistic point
 contact with a characteristic size $d$ causes a reduction of the
 contact conductance \cite{K84b}
 \begin{equation}\label{Delta_G_i}
 \delta G_i \approx - {{\mid V_0^{(i)} \mid }^2 \over
 {\varepsilon}_F^2}\left({{\lambda_F}^2 \over d^2}\right) \, G ,
 \end{equation}
 where $\lambda_F$ is the Fermi wavelength, and $G \approx
 (e^2/h)(d^2/\lambda_F^2)$ is the Sharvin conductance. Making use of
 the Eq. (\ref{E_eli}) one obtains $\delta G_{i} \approx
 (e^2/h)(E_{el,i}/\varepsilon_F)$.  Therefore, if the defect asymmetry
 is completely controlled by the conduction electrons, the conductance,
 which corresponds to the defect in its lower state (larger absolute
 value of the electron contribution to the defect energy) is smaller
 than that for the defect in the upper state.
 
 \subsection{Interference contribution to the renormalization}
 
 Let us consider now the electronic interference.
 For adiabatic effects different states of ``active'' defects,
 fluctuators, can be considered as independent realizations. Thus we
 can choose some configuration of some active defect as a ``reference''
 scatterer $i$ and consider its properties in the presence of
 ``background'' scatterers.  We will analyze both the
 interference contribution to conductance due to the defect $i$ and the
 ``mesoscopic'' renormalization of the energy of this defect, which is
 equal to a change of the electron system energy due to a presence of
 interference pattern involving defect $i$.
 
 The ballistic point contact contains finite number of scatterers.
 Therefore the interference contribution to the contact conductance is
 provided by a {\it local} interference (involving
 trajectories with small number of scatterers)
 rather than by a {\it global} one (which leads to
 well-known UCF \cite{Magfield}).  This ``local'' interference
 contribution has been to some extent
 analyzed in Ref.\cite{Galp91}. In what follows for simplicity we
 will restrict ourselves mainly to the interference patterns involving
 only pairs of scatterers. However, as it is shown in Appendix 1a,
 our results can be generalized for the case of arbitrary
 number of scatterers.

 As it is shown in Appendix 1a [Eq.(\ref{Appres})], the contribution to the
 conductance due to a pair of scatterers, namely ``reference''
 scatterer $i$ and the ``background'' $m$, is:
 \begin{eqnarray}\label{G_im}
 \delta G_{im} &=& A_{G,im}\; \xi_{im}, \nonumber \\
 A_{G,im} &\approx & {e^2 \over h} {|V_0|^2 \over \varepsilon_F^2}
 \left({\lambda_F \over R_{im}}\right)^2 .
 \end{eqnarray}
 Here $\xi_{im} = \cos (2 k_F R_{im})$ and ${\bf R}_{im} \equiv {\bf
 R}_i - {\bf R}_m$ is the vector, which connects two scatterers.  For
  simplicity we assumed that the scattering potentials for all
 scatterers, both ``active'' and ``passive'' ones, depend only on the
 coordinate ${\bf R}_i$ of the scatterer: $V_{i{\bf k k'}} = V_0
 \exp[i{\bf (k - k')R}_i]$. 
 
 Let us now find a contribution to the energy of ``reference''
 scatterer $i$, which represents one state of some fluctuator,
 due to its being involved in electron interference along with another
 scatterer $m$. Following the scheme implied by Eq. (\ref{E_eli}), it
 is given by a renormalization of electron system energy due to this
 pair of scatterers. In second-order perturbation theory approximation
 we obtain, collecting all terms proportional to $ V_{i{\bf k k'}}
 V_{m{\bf k k'}}^*$ :
 \begin{equation}\label{E_im1}
  E_{el,im} =
 \mbox{ \bf Re} \sum_{{\bf k}, {\bf k}'} |V_0|^2
 \exp \left[i({\bf k} - {\bf k}'){\bf R}_{im}\right] \;
 {f(\varepsilon_{\bf k}) - f(\varepsilon_{{\bf k}'}) \over
 \varepsilon_{\bf k} - \varepsilon_{{\bf k}'}}.
 \end{equation}
 A straightforward calculation for a spherical Fermi surface
 and zero temperatures gives [see Appendix 1a, Eq. (\ref{Appres})]:
 \begin{equation}\label{E_im2}
 E_{el,im} = A_{E,im}\; \xi_{im},  \mbox{\hskip1cm}
 A_{E,im} \approx {{\mid V_0 \mid}^2
 \over {\varepsilon }_F }{\lambda_F^3 \over R_{im}^3} \, .
 \end{equation}
The obtained renormalization is due to interaction of defect $m$ with the
Friedel oscillation of electron density originated by the defect $i$. 
 
 Both $\delta G_{im}$, Eq. (\ref{G_im}), and $ E_{el,im}$,
 Eq. (\ref{E_im2}), are proportional to the {\it same} phase factor
 $ \xi_{im} = \cos (2 k_F R_{im})$. Correspondingly,
 \begin{equation}\label{scal}
 {\left[ \delta G_{im}/(e^2/h) \right] \over (\delta E_{el,im}/
 \varepsilon_F)} \sim {\lambda_F \over R_{im}} > 0
 \end{equation}
  and thus, in analogy with the
 ``classical'' effect, the larger is the energy of a configuration, the
 larger is the contact conductance. As it can be shown
 (see Appendix 1a) the proportionality to the same phase factor and,
therefore, Eq.\ref{scal}, holds not only for pairs of scatterers but for
 arbitrary number of scatterers as well.

 To estimate a total interference contribution due to the defect $i$
 both to the conductance, $\delta G_i$, and to the defect energy,
 $E_{el,i}$, one must sum over all ``background'' scatterers $m$.
 Due to a random distribution of ${\bf R}_{im}$ this results in some
 mesoscopic fluctuations for both quantities with respect to
 realizations of the system. However the fact that both quantities are
 linearly related to the same set of random factors $\xi_{im}$ leads
 to the correlation between them, namely 
 \begin{equation} \label{corr}
 \left<\delta G_i \, E_{el,i}\right> = C \, \overline{\delta G} \,
 \overline{E_{el}},
 \end{equation} 
 where $\overline{ x } \equiv \sqrt {\left< x^2 \right>}$,
 $C \approx 1$, and $\left< \; \right>$ denotes the ensemble average.
 More detailed argumentation of this is given in Appendix 2.
 For a given value of $E_{el,i} = E$ one has
 \begin{equation}\label{<G_i>}
 \left< \delta G_i \right>_E = C\overline{\delta G}
 {\mbox{\hskip0.5cm} E \mbox{\hskip0.5cm} \over \overline{E_{el}}} .
 \end{equation}
 Keeping in mind factors $1/R_{im}^2$ (for the conductance) and
 $1/R_{im}^3$ (for the energy) one may suggest
 the main contribution to both these
 quantities to stand from the nearest neighbors.
 In this case  both energy renormalization
 and contribution to resistance would be  related to a few neigbouring
 defects and thus could be estimated
 by Eqs.\ref{G_im}, \ref{E_im2} with $R_{im}$
 of the order of most probable interdefect distance $ N_i^{-1/3}$
 (where $N_i$ is the background defect concentration). Note that if
 we took an average over all possible realizations of
 the background scatterers we would met the problem with
 singular behavior of the averaged quantities when $R_{im} \rightarrow
 0$. It means that the average is mainly controlled by
 (rare) realizations corresponding to very close neiubouring background
 defect, and one has $R_{im} > k_F^{-1}$, that 
 would give $|\delta G| \sim e^2/h$,
 $E_{el,i} \sim E_F/(k_Fl)^{1/2}$. However, in the
 case of small ballistic contact we deal with some given contact
 realization, so that one deals with the 
 most probable rather than with average quantity (compare with \cite{Alt95}).
 
 For the ballistic point contact the number of defects in the contact is
 small, so that the dominant contribution is expected to be
 from the trajectories,
 which involve a boundary of the contact. Assuming the contact to be a
 short channel with a length $\approx d$, this boundary may be
 considered as an array of scatterers at a distances $\approx d $ from
 the defect $i$ with a total number $\approx (d/\lambda_F)^2$.
 For this case
 \begin{equation}\label{<G>}
 \overline{\delta G} \approx
 \left({\lambda_F \over d}\right){e^2  \over h}, \mbox{\hskip0.5cm}
 \overline{ E_{el,i}} \approx \varepsilon_F \left({\lambda_F
 \over d}\right)^2 .
 \end{equation}
 
 Taking values of $d$ typical for nanofabricated ballistic point
 contacts,  $d\approx 5-10 nm$ one gets $|\delta G| \approx (0.05 - 0.1)
e^2/h$, $E_{el,i} \approx 30 - 100 K$.
 
  The mesoscopic
 interference renormalization has some special features as compared
 with the ``classical'' one. First, mesoscopic disorder lifts energy
 degeneracy of the defect states, which have different spatial
 positions. Thus it causes a formation of fluctuators from otherwise
 symmetric defect configurations (to say, interstitials, which have
 symmetrical lattice positions).
 
 Then, in this case one expects the temperature and bias behavior of
 the resistance to depend on an external magnetic field, which affects
 the electron interference (see e.g. Ref. \cite{Magfield}).
 
 In addition, the interference contribution both to the defect energy
 and to the conductance depends on the electron distribution. The
 finite applied bias makes this strongly nonequilibrium, which at
 high enough biases causes a ``direct'' effect of bias on both
 quantities.
 
 The physical picture of the mesoscopic renormalization is much richer
 than that provided by ``classical'' one, and it is this mechanism
 that we will concentrate on in the rest of the paper.
 
 \section{Zero-bias and zero-temperature resistance anomalies}
 
 Let us consider defect $i$ which occupies either of the {\it two}
 neighboring positions, 1 and 2, with close energies. For simplicity
 we will assume that the energy asymmetry of these defect states is
 completely determined by the electron renormalization.  This object
 is a sort of two-level fluctuator originated from the {\it
 electronic} disorder rather than from the lattice one.  It is
 important, that due to the correlation discussed above, the upper
 state of such fluctuator, which corresponds to a defect position with
 higher energy, gives a smaller contribution to the contact resistance.  A
 conductance increase, which accompanies a transition from the lower to the
upper
 level of such fluctuator, $\delta G_{(i)} = \delta G_{(i)1}- \delta
G_{(i)2}$ is,
 according to Eq. (\ref{<G_i>}), scaled with the energy asymmetry
 $E_{(i)} = E_{el,(i)1} - E_{el,(i)2}$. Here index $i$ now denotes the
 fluctuator.
 
 A summation over all fluctuators gives their total
 contribution to the average contact conductance:
 \begin{equation}\label{Delta_G}
 \Delta G = \sum_i \delta G_{(i)} n_{(i)} =
 \int \left< \delta G_{(i)} \right>_E P(E)\, n(E) \, dE.
 \end{equation}
 Here $n_{(i)} = n(E_{(i)})$ is the $i$-th fluctuator upper level
 occupation number and $P(E)$ a density of states given by statistical
 properties of $E_{el,i}$. For the mesoscopic system it is
 reasonable to take the values of $E_{el,i}$ for the neighboring
 defect positions as statistically independent. In this case $P(E)$ is
 approximately constant at small $E \ll \overline{E_{el}}$.
 Making use of Eq.(\ref{<G_i>}) and of
 the expression $n(E) = [1 + \exp (E/T)]^{-1}$ one has at small
 temperatures $T \ll \overline{ E_{el}}$ the conductance enhancement
 \begin{equation}\label{beta}
\Delta G \propto T^{\beta},
\end{equation} 
where $\beta =2$.
 
 For some defects, like light interstitials or some defect complexes,
 the probabilities for defect hopping between spatially symmetric
 positions are relatively high \cite{Hultman}. For these ``delocalized''
 defects the effect of electronic disorder provides a many-site
 ``potential relief'' instead of two-site fluctuator picture.Assuming that any site can be occupied by only one defect one
deals with "Fermi-type" statistics; so at $T \rightarrow 0$ the
sites with lowest energies are occupied by the mobile defects
while those with energies higher than the "Fermi level" are free.
At finite temperature Eq.\ref{Delta_G} can be applied where the
site occupation number has again the form $n(E) = [1 + \exp
(E/T)]^{-1}$ if one takes the "Fermi level" as the origin for
the energy $E$. In this case  the total number of available sites is
much larger than the number of defects $N$, and for
finite temperatures the "Boltzmann-type" statistics holds rather than
the "Fermi-type":
\begin{equation}\label{Boltz}
n(E) =  {N \, \exp (- E / T)\over \int P(E)\exp (-E
/ T){\rm d}E}
\end{equation}
In this case a change of $T$ does not affect the number of rearranged
defects (because any of them can change its energy) and leads only to
a change of the average defect energy. As a result,  in this case we have
in Eq.(\ref{beta}) $\beta = 1$ independently of the form of the density
of states $P(E)$.

 Let us turn now to the effect of finite bias $eV \gg T$.  For TLS it
 was first considered in Refs. \cite{K84b,K86}. It was shown that for
 low-energy TLS with small enough energy splitting $E$, for which the
 coupling with electrons dominates \cite{Black81}, the TLS occupation
 numbers are sensitive to the electron distribution. For the contact
 region this is strongly nonequilibrium and for a central point of
 symmetric contact has a form
 \begin{equation}\label{f(V)}
 f({\bf k}) = \theta (k_x) f_0(\varepsilon_{\bf k} + eV/2) + \theta
 (-k_x) f_0 (\varepsilon_{\bf k} - eV/2),
 \end{equation}
 where $OX$ is the main axis of the contact and $\theta (x)$ the
 theta function.  The ``energy width'' of this distribution, $eV$,
 plays a role of the effective temperature. In particular, the upper
 levels of the TLS are empty if $eV < E$, while for $eV \gg E$ the
 occupation numbers of TLS levels are almost equal \cite{K84b,K86} and
 $n(E) \approx (1/2)[1 - (E/eV)]$.
 
 For larger $E$ the coupling with phonons becomes important
 \cite{Black81,KR93}
 due to rapid increase of density of states for the actual phonons
 with $E$ increase.  For the two-state case the fluctuator relaxation
 rate due to electron-assisted tunneling is \cite{K84b,Black81} $
 W_{el}(E,V) \approx (|V^-_0|/\varepsilon_F)^2 [(eV - E)/\hbar]{\cal
 T}$, where $V^-_0 = (V^{(2)}_0 - V^{(1)}_0)/2$. For the
 phonon-assisted process \cite{Galp89} $W_{ph}(E,T) \approx (\Lambda^2
 E^2 / {\cal E} \Theta_D^3) (E/\hbar) {\rm cth}(E/2T){\cal T}$. Here
 $\Lambda$ is the fluctuator deformational potential, ${\cal E}$ and
 $\Theta_D$ are the atomic and Debye energies, respectively.  ${\cal
 T}=\exp(-\lambda)$, where $\lambda$ is the tunneling constant.
 
 Let us define the characteristic energy $E^{*}$ for which
 $W_{el}(E^*,(eV - E^*) \approx E^*) = W_{ph}(E^*,T=0)$; for the
 reasonable values of parameters (see e.g. Ref.\cite{Black81}) $E^*$ is
 expected to be $\approx 1 - 3 K$. For $eV \gg E^{*} $ a probability
 of electron-assisted tunneling to the upper level $W_{el}$ exceeds
 the probability of phonon-assisted decay of the upper level $ W_{ph}$
 up to some threshold energy $E = E_{th} = E^{*}(eV/E^{*})^{1/3}$ at
 which an increase of the electron phase volume with bias ($\propto
 eV$) is compensated by a corresponding increase of phonon phase
 volume ($\propto E^3$). For the crude estimates let us take the
 occupation numbers $n(E) \propto \theta (E_{th} - E)$.
 Now making use of Eq. (\ref{Delta_G}) and assuming 
 density of states $P(E)$ constant, one obtains the following
interpolation formula for the interference contribution to the conductance:
 \begin{equation}\label{interp}
 \Delta G \propto \left[ E^* \left({eV \over E^{*}}\right)^{1/3} + T
 \right]^{\beta}
 \end{equation}
with  $\beta=2$.

 The same
considerations can be applied for the case of "delocalized" defects. 
Although the probabilities $W_{el}$ and $W_{ph}$ for this case can
differ from ones for the two-level fluctuators, the scaling
$W_{el}/W_{ph} \sim eV/E^3$ (relation between relevant 
electron and phonon phase volumes)
 holds for ($eV \gg E \geq T$) and thus Eq.\ref{interp}
is valid, but with $\beta = 1$.
 
\section{Direct effect of the applied bias on the fluctuator parameters}

When a large enough bias $V$ is applied to the point contact one
should take into account the non-equilibrium electron distribution in
course of estimates Eqs.(\ref{G_im}), (\ref{E_im2}), (\ref{<G_i>}).
For the distribution given by Eq.(\ref{f(V)}) one obtains (see Appendix 1b)
for
$\Delta E_{el,im}$ the phase factor
\begin{equation}\label{VE} 
\Delta E_{el,im} \propto \cos [2 k_FR_{im} + \phi(V,R_{im},k_F)]
\cos (2\Delta k R_{im}), 
\end{equation}
and for $\delta G_{im}$ a factor  
\widetext
\begin{eqnarray}\label{VG}
 \delta G_{im}   \propto  \cos [2k_FR_{im} + \phi (V,R_{im},k_F)]& &
\cos (2\Delta k R_{im}) \nonumber
\\
 & + & {\partial \phi \over \partial V}{1 \over 2 R(\partial \Delta k /
 \partial V)} \sin [2k_FR_{im} + \phi(V,R_{im},k_F)]\sin
(2\Delta k R_{im}),
\end{eqnarray}
\narrowtext
where  $\Delta k \equiv k_F
eV/\varepsilon_F $.  This is the ``direct'' effect of bias on the
fluctuators parameters in addition to tuning  of fluctuator 
level occupation numbers. 

As seen, the first term in Eq.\ref{VG} is correlated with the phase
factor of Eq.\ref{VE}, while the second is not and, thus, will sum 
out. As for the correlated cosine terms, the effect of bias initially ( at 
$\Delta k R \ll 1$) leads to their decrease due to a decrease of the
corresponding  cosine factors, while for
$\Delta k R \gg 1$  (when the factors are random with respect
to parameter $R_{im}$ ) they are
suppressed due to additional (with respect to the case $V = 0$)
averaging over $R_{im}$: 
$$
(\overline{\Delta E_{el}},\overline{\Delta G}) 
\propto (\Delta k R)^{-1/2}.
$$
Actually we deal here with the well-known energy
averaging effect suppressing any mesoscopic phenomena.

It is important to note that these effects can lead to the
resistance anomalies even if the defect structure is not rearranged
in course of the bias application; the only condition is
that the defects occupy the positions with the lowest
energies available and thus with largest mesoscopic contribution to
resistance $\overline{\delta G}_i$. 
For these defect 
configurations one gets in average $\overline{\delta G} < 0$. Total
interference contribution to the conductance due to  $N$ defects is:
\begin{equation}
\Delta G \sim N {\cal F}(V) \overline{\delta G},  
\end{equation}
where ${\cal F} \sim \cos (\Delta k R)$ for $\Delta k R < 1$, and 
${\cal F} \sim (\Delta k R)^{-1/2}$ for $ \Delta k R \gg
1$.  
The result we obtained is that  the bias increase leads to a systematic
conductance increase. It is interesting that in combination with the
effects discussed for relatively small $V$ --- occupation of
states with higher energies, this "direct" effect can form
the configurations with {\it smaller} resistances than
available for a simple temperature increase. Indeed,
it can suppress (negative in average) 
mesoscopic contribution to conductance due to
configurations with large enough energy 
gap between the available realizations which can not be rearranged
at relatively small temperatures when the phonon contribution to
resistance (obviously masking the effects in question) is still 
small. Note that as we have seen above the bias values allowing the
same occupation states of the defects and at the same time the same
efficiency of electron-phonon processes as in equilibrium state
with a temperature $T$ is scaled with $T$ as 
$eV \approx T (T/E^{*})^3$. Thus the energy averaging effects can
become to be pronounced for large-gap configurations when the
filling of the upper level is still negligible.

Certainly, the temperature
increase can also lead to the energy averaging, but the necessary
temperatures are too large and correspond to a significant phonon
contribution to resistance.

 \section{Discussion}
 In this section  we would like to make several remarks about the
limitations and possible complications of our model. 

 First, until now we considered the defect energy density of states as
 constant. The limitation of the defect energy band leads to a
 saturation of $\Delta G(T)$ and $\Delta G(V)$ dependencies at $T >
 T_{sat}$ and $V > V_{sat}$, respectively.  These quantities scale as
 $eV_{sat} \approx T_{sat} (T_{sat}/E^{*})^3$, and $\Delta G_{sat}
 \approx N_f \overline{\delta G}$, where $N_f$ is a
 total number of fluctuators. As for the estimate for $T_{sat}$,
 taking $\varepsilon_F \approx 10^4 - 10^5$K and $(d/\lambda_F)\approx
 50$ and making use of Eq. (\ref{<G>}) we obtain $T_{sat} \approx
 \overline{E_{el,i}} \approx 4 - 40$K. Note that the ``saturation'' value
 $\Delta G_{sat}$ corresponds to a random realization of different
 interference patterns involving the fluctuators, while the
 values of $\Delta G$ at lower temperatures correspond to
 preferable occupation of larger resistance states and thus are
 systematically smaller than typical for mesoscopic disorder.
 
  Second, it is important that the picture discussed is sensitive to the
 external magnetic field. In particular, it is known that in
 homogeneous diffusive conductors the interference particle-particle
 channel is 
 suppressed in the strong enough magnetic field $H > \Phi_0/L_c^2$,
 where $\Phi_0$ is the quantum of the magnetic flux and $L_c$ is a
 coherence length, instead of which in our case we should use a contact size
 $d$. This suppression reduces a magnitude of mesoscopic fluctuations
 nearly twice \cite{Alt89,Magfield}. Point contact is a strongly
 inhomogeneous system and the main contribution to the mesoscopic
 fluctuations is due to local interference. However, despite the fact,
 that the effect of magnetic field implies a contribution of
 configurations which involve more than two scatterers, this
 contribution is relatively large due to rather high probability of
 the boundary scattering and lead to decrease of $\Delta G$ with field
 increase.

 Another important feature is related to a coupling between different
``active'' defects, $i$ and $j$, due to a dependence of the defect
$i$ energy on a position of defect $j$. This dependence is given by
Eq.(\ref{E_im2}).  For large enough concentration of ``active''
defects one may expect a formation of self-organized aggregates in
the defect system (of spin-glass type). Indeed, the defect positions
corresponding to maxima of the Friedel oscillations originated
by the  other defects become energetically preferrable, which introduces
some "ordering" into the defect system. Thus a formation of
"coherently scattering" aggregates can be possible leading to
a significant enhancement of the interference contribution to
resistance. The increase of bias
is expected first to suppress  this contribution to the  resistance in a
way similar to discussed above. On the other hand, at higher biases 
the direct bias-induced
decrease of ``coupling potentials'' $E_{el,ij}$ (see Eq. (\ref{VE}))
 can destroy such
aggregates, which can lead to sharp resistance changes.
 
 Finally, it is instructive to compare results of our model with the two-channel
Kondo model (TCKM)
 \cite{Zawadovskii,Ralph}, which also predicts zero-bias resistance anomalies
 of a non-magnetic nature. Despite these two models are based on completely 
different physical background, they predict qualitatively similar resistance 
behavior at  low $T$ and $V$: negative temperature and bias resistance 
coefficients affected by a magnetic field.  However, quantitative predictions
of the two models differ. Our model does not
 predict a singular $T$-behavior at $T \rightarrow 0$ - in
 contrast to TCKM.  As for bias dependence, our  model predicts singular
 behavior, $V^{2/3}$ for biases $V \gtrsim 1-3$mV (see Eq.(\ref{interp})) and
saturation  at smaller biases. This saturation can imitate the "restoration of
the Fermi liquid behavior" predicted by TCKM.  On the other hand,  TCKM, being
related to
 non-adiabatic effect, is relevant to 
 the fluctuators  of a rather special type (with a small asymmetry and large
 tunneling probability), while  our predictions 
 hold for any sort of mobile defect. Our model also predicts 
 special features at higher temperatures and biases; in particular,  the
saturation of the zero-bias anomaly at large $V$ and $T$, and a principal
possibility to reach larger values of conductance in course of bias increase
with respect to ones available for temperature increase. 
 
 \section{Conclusions}

To conclude, we predict a new mechanism of zero-bias resistance
anomalies in metallic point contacts based on a found correlation between
energies of defects with internal degree of freedom and
their contributions to resistance. The correlation
lifts to some extent the "random" character of mesoscopic disorder and 
breaks the symmetry of the defect states with respect to the
signs of the mesoscopic contribution to resistance.
 We  suggest a model of a fluctuator related
to a purely electronic disorder, which provides a new insight into the 
 nature of fluctuators in the perfect point contacts.

 We are indebted to Yu.M.Galperin and H.R.Schober for reading the
 manuscript and extremely valuable remarks, and to B.L.Altshuler for
 fruitful discussions. V.I.K. gratefully
 acknowledges the ``Nederlandse Organisatie voor Wetenschappelijk
 Onderzoek'' for a visitors grant, while A.M.R. acknowledges the
 financial support of German Ministry of Technology under the
 German-Russian cooperation agreement and the hospitality of
 Forschungszentrum Julich, where part of this work was done. Partial support by 
Russian Foundation for Fundamental Research (Grant N 95-02-04109-1) is also 
gratefully acknowledged. 
 
 \section{Appendix 1}
 \subsection{Calculation of the mesoscopic contribution to
the conductance and to the defect energy at small applied biases. }

To calculate mesoscopic contributions both to conductance
and to energy of electron system
due to the presence of some finite number of scatterers we
will make use of the  "wave-optics"
approach \cite{kozub94}. The approach is based on perturbation theory
in real space. Let us consider an electron, the wave function of
which initially is a plane wave with the wave vector $\bf k$.
After $n$ successive scattering events involving
scatterers $1, ..., n$ the electron wave function becomes:
\begin{eqnarray}\label{scatstate}
|1, &..., &n> \equiv \psi_{1,..., n}({\bf r}) = {f^n \over |{\bf r} -
{\bf R}_n|\cdot ... \cdot
|{\bf R}_2 -
{\bf R}_1|} \nonumber\\
& \times & e^{i{\bf k R}_1}\exp \left(ik [|{\bf r} - {\bf R}_n| +
... +|{\bf R}_2 - {\bf R}_1|] \right)
\end{eqnarray}
where ${\bf R}_1, ... {\bf R}_n$ are the positions of the scatterers. For
the short-range scatterers the scattering amplitude $f$ in the Born
approximation takes the form:
\begin{equation}
f = -{m \over 2 \pi \hbar^2}\int {\rm d}^3 r V({\bf r})
\end{equation}
and is assumed 
the same for all scatterers. Contribution of
the scatterers to the conductance is determined by their backscattering
efficiency.
The interference contribution to
the backscattering current
due to trajectories involving scatterers $1, ... ,n$
and $1', ..., n'$ is
\begin{equation}\label{curr}
\delta {\bf j} = {ie \hbar \over 2m}(<1',...,n'| \nabla |1, ..., n> +
c.c.).
\end{equation}
To obtain a contribution to the conductance one should integrate
this equation over $\bf r$ within some
reference plane remote from the scatterers system. It is important
that
only position of last scatterers $n$ and $n'$ are relevant for this
integration and one deals with a factor
$$ \int{\rm d}^2 \rho \exp \left[ ik ( |{\bf r} - {\bf R}_n| -
|{\bf r} - {\bf R}_{n'}|)\right]. $$
Here $\rho$ is a projection of $\bf r$ on the plane in question.
Taking the plane to be normal to ${\bf R}_n - {\bf R}_{n'}$
and
expanding the exponent as $[...] \sim |{\bf R}_n - {\bf R}_{n'}|
+ \rho^2 |{\bf R}_n - {\bf R}_{n'}|/r^2$ one gets the result of the
integration in the form:
$$ {1 \over ik|{\bf R}_n - {\bf R}_{n'}|}\exp (ik|{\bf R}_n - {\bf
R}_{n'}|). $$
The next step is the 
integration over $\bf k$ directions. In its turn, this integration
is relevant
only to co-ordinates of the "first" scatterers in the chains that is to
${\bf R}_1$ and ${\bf R}_{1'}$, which  enters the exponential factor
$\exp [{\bf k}({\bf R}_1 - {\bf R}_{1'})]$. Correspondingly, the
integration over $\cos \theta$ where $\theta = \angle ({\bf k},
{\bf R}_1 - {\bf R}_{1'})$ gives the factor
$$
 {1 \over ik |{\bf R}_1 - {\bf R}_{1'}|} \exp (ik |{\bf R}_1 - {\bf
R}_{1'}|).
$$
Finally one arrives at the following estimate for the mesoscopic contribution
to the conductance:
\begin{eqnarray}\label{mescond}
{\delta G \over G} & \sim & {f^{n + n'} \over k^2a^2 R^{n + n'}} \nonumber \\
& \times& \cos
\left[ \varphi(n,n',n'-1, ...,1',1; n', n,n-1, ..., 1)\right]
\end{eqnarray}
where the phase $\varphi$ is
\widetext
\begin{eqnarray}\label{phase}
  \varphi(n, n',n'-1, ...,1',1;  n,n-1, ..., 1) & & \nonumber\\
  = k(|{\bf R}_{1'}  - {\bf R}_1|  +  |{\bf R}_{n'} - {\bf R}_n| &+& |{\bf R}_{n'} - {\bf R}_{n'-1}|  +  ... + |{\bf R}_{2'} - {\bf R}_{1'}| 
- |{\bf R}_n - {\bf R}_{n-1}| - ... - |{\bf R}_2 - {\bf R}_1|).
\end{eqnarray}
Here $R$ is a typical interscatterer distance within the chains
while $d$ is the contact size appearing as a result of normalization of
the backscattering efficiency on the incident electron flow.

The picture discussed can be interpreted as a
contribution to scattering due to presence of
the scatterer $n$ affecting the superposition of
states formed by successive scattering by chains $1, ..., n-1$ and
$1', ..., n'$. The phase $\varphi$ which after the integration over
$\bf k$ directions is the phase difference for the paths
$n,n',n'-1, ..., 1',1$ and $n, n-1, ... ,1$, correspondingly.
One should also note that in course of derivation of
Eq.\ref{mescond} we have taken into account that only those electrons
with energies close to the Fermi energy contribute to the conductance, 
and  used $k = k_F$ in Eq.(\ref{mescond}).

Now let us estimate the mesoscopic contribution to the electron energy
due to the presence of the same system of the scatterers finally
affecting the electron state in the position of  scatterer $n$.
In the lowest approximation one has
\begin{equation}\label{ren}
\delta E_{el} = <1', ..., n'|V({\bf r} - {\bf R}_n)|1, ..., n-1 >
\end{equation}
where $V$ is a scattering potential
assumed to be short-range: $V = V_0 \delta (r)$.
As a result of averaging over the
direction of $\bf k$ we obtain
\begin{eqnarray}\label{ren2}
\delta E_{el} & \approx &  - V_0 {f^{n + n'-1} \over k R^{n + n' + 1}} \nonumber \\
 & \times & \sin [k
\varphi (n, n',n'-1, ...,1',1;  n,n-1, ... 1)]
\end{eqnarray}
It is important that due to the same structure of the
expression for $\delta E_{el}$ as of one for the $\delta G$
the phase difference for the interference
pattern $\varphi$ is exactly the same.

In order to obtain the interference correction to the energy
of the whole electron system we should sum Eq.\ref{ren2}
over all
occupied electronic
states.
For $T = 0$ one
has
\widetext
\begin{eqnarray}\label{ren3}
\delta E_{el} \sim  - {V_0  \over \pi^2} {f^{n + n'-1} \over  R^{n + n' + 1}}
 \int_0^{k_F}k{\rm d}k \sin [k
\varphi (n, n',n'-1, ...,1',1;  n,n-1, ... 1)] = \nonumber\\
\nonumber\\
= - {V_0 \over \pi^{2}} {f^{n + n'-1} k_F\over  R^{n + n' + 2}(n + n')}
\cos [\varphi (n, n',n'-1, ...,1',1;  n,n-1, ... 1)]
\end{eqnarray}
\narrowtext
Thus we conclude that the mesoscopic contributions
of the same system of scatterers to
the conductance and to electron energy renormalization
depend on the {\it same} phase factor and, therefore, {\it are
correlated}.

For the simplest case of 2 scatterers (positioned in  ${\bf R}_1$ and
${\bf R}_{1'}$) site ${\bf R}_1$ plays at the same time a role
of the site ${\bf R}_n$. The phase factor in this case is
$$
\cos (k|{\bf R}_{1'} - {\bf R}_1| + k|{\bf R}_{1'} -
{\bf R}_1|) = \cos (2k|{\bf R}_{1'} - {\bf R}_1|) 
$$
and, correspondingly,
\begin{equation} \label{Appres}
 {\delta G \over G} \sim {f^2 \over a^2 k_F^2 R^2}
\cos (2 k R), \,\,\,\,\, \delta E_{el} \sim - {fV_0k_F \over R^3}
\cos (2kR) 
\end{equation}
which gives   Eqs.(\ref{G_im}) and (\ref{E_im2}). 

%We would like to note,
% that as for the renormalization, one may rewrite the result obtained
%in more standart momentum representation, noting that
%$$\int {\rm d}^3 r {e^{ikr} \over r}\exp (i{\bf q r}) =
%{1 \over (k^2 - q^2) } $$
%Thus for the quadratic spectrum one comes to the standart result
%$$ \Delta E_{el} = \sum_{\bf k,q} |V_{kq}|^2{ F(\varepsilon_k) \over
%\varepsilon_k - \varepsilon_q} = \sum_{\bf k,q} |V_{kq}|^2
%{F(\varepsilon_k) - F(\varepsilon_q) \over
%\varepsilon_k - \varepsilon_q}  $$

\subsection{Direct effect of bias on the interference
contributions}

In this subsection we will study the  "direct" effect of
bias on the renormalization of the fluctuator energy and on the
interference contribution to the conductance  for the simplest case of 2
scatterers. 

We start with deriving 
$E_{el}$. For the step-like electron distribution
given by Eq.\ref{f(V)} one obtains for the integral over $k$ (instead of
Eq. (\ref{ren3})):
\widetext
\begin{eqnarray}
{1 \over 2}\int_{k_F - \Delta k}^{k_F + \Delta k}{\rm d}k\cdot
\sin (2kR + \phi (V,R,k_F)) +  \int_0^{k_F - \Delta k}{\rm d}k\cdot
\sin (2kR + \phi (V,R,k_F)) = \nonumber \\
\int_0^{k_F + \Delta k}{\rm d}k\cdot k \sin (2kR + \phi (V, R,k_F)) -
{1 \over 2}\int_{k_F - \Delta k}^{k_F + \Delta k}{\rm d}k \cdot k
\sin (2kR + \phi (V,R,k_F))
\end{eqnarray}
Here $\Delta k = eV/\hbar v_F$. 
An additional $V$-dependent phase $\phi$ is related to a dependence of
$k$ on coordinate due to a presence of electric field
($(\hbar k )^2/2m + \varphi ({\bf r}) = \varepsilon = const$).
Calculation of the integral gives the phase factor 
\begin{equation}\label{VEF}
\cos [2(k_F + \Delta k)R+ \phi(V,R,k_F) ] + \sin (2k_FR + \phi) 
\sin (2 \Delta k R ) =
\cos [2k_FR + \phi (V,R,k_F)] \cos (2 \Delta k R)
\end{equation}
instead of the factor $\cos (2 k_FR)$obtained for $V \rightarrow 0$.

In the same way we estimate the contribution to conductance at $T =0$:
\widetext
\begin{eqnarray} \label{lala}
\delta I & \propto & \int_{k_F - \Delta k}^{k_F + \Delta k} \cos
(2 k R + \phi(V,R,k_F)), \nonumber 
\\
\delta G & = &{{\rm d} I \over {\rm d} V} \propto
{\partial \Delta k \over \partial V} 
\left[ \cos (2(k_F + \Delta k)R + \phi(V,R,k_F)) + \cos
[(2(k_F - \Delta k)R] + \phi(V,R,k_F)) \right] \nonumber 
\\
& - & {\partial \phi \over
\partial V} \int_{k_F - \Delta k}^{k_F + \Delta k} \sin
(2 k R + \phi(V,R,k_F)) \nonumber \\
& \sim  &{\partial \Delta k \over \partial V}
2 \cos (2k_FR + \phi (V,R,k_F))\cos (2\Delta k R) 
+{\partial \phi \over \partial V}{1 \over 2 R}
2\sin (2k_FR + \phi(V,R,k_F))\sin
(2\Delta k R)
\end{eqnarray}
\narrowtext
Taking into account that $\phi \sim \Delta k R\cdot R/a$, one sees
that  the second and the first terms  in r.h.s. of Eq.(\ref{lala}) are of
the same order provided that  $R/a \approx 1$. However the first
term completely correlates with the corresponding phase factor  
for the energy renormalization, Eq. (\ref{VEF}), while the second term does
not.

 \section{Appendix 2}
 
 For each "active" defect $i$ the interference contribution to the
 conductance, $G_i$, as well as to  the energy, $E_{el,i}$ contains
 summation over
 "background" scatterers $m$. Contribution of each scatterer
 $m$ gives some phase factor $\xi ({\bf R}_{im})\equiv
 \xi_{im}=\xi_{mi}$ which depends 
 on the distance of the scatterer $m$ from the defect. Hence, one can rewrite     
the expressions for $\delta G_i$ and $E_{el,i}$ in a form:
 \begin{eqnarray}
 \delta G_i & \equiv & \sum_m G_{im} \xi_{im} \equiv ({\bf G}_i, \vec
 \xi_i), \\
 E_{el,i} & \equiv & \sum_m E_{im} \xi_{im} \equiv
 ({\bf E}_i, \vec \xi_i)
 \end{eqnarray}
 Here we have introduced some "vector space", where  vector
 $\vec{ \xi}_i$ contains the set of the corresponding phase factors,
 and  vectors ${\bf E}_i$ and ${\bf G}_i$ contain
 the sets of the prefactors (given by Eqs. (3) and (5),
 correspondingly).  
 For the ensemble of defects $i$ the vector $\vec{\xi}_i$ should
 be considered 
 as random, while all components of $\bf E_i$ and $\bf G_i$ are positive.
 We may rewrite the vectors ${\bf E}_i$, ${\bf G}_i$ as
 \begin{equation}
 {\bf E}_i = {{\bf E}_i \over {\bar E_i}} {\bar E_i}, \,\,\,\,\,
 {\bf G}_i = {{\bf G}_i \over {\bar G_i}}{\bar  G_i}
 \end{equation}
 where we have introduced the "norms" of the vectors ${\bf E}_i$ and
 ${\bf G}_i$.  The correlator $\left<\xi_m , \xi_n \right> = \gamma
 \delta_{m,n}$ (where for the cosine phase factors $\gamma = 1/2$),
 and we obtain

 \begin{equation}
 <\delta G_i, E_{el,i}> = < ({\bf G}_i, {\bf \xi}_i)({\bf E}_i, {\bf
 \xi}_i)> = \gamma
 \bar E_i \bar G_i ({{\bf E}_i \over \bar E_i}, {{\bf G}_i \over
 \bar G_i})
 \end{equation}
 The scalar product of the normalized
 positively-defined
 vectors in the brackets is of the order of unity, and one comes to the
 estimate for the average,  Eq.\ref{corr}, $<...> =
 C {\bar \delta G}{\bar E_{el}}$ (according to definitions given in
 front of Eq.7, ${\bar E} = \bar {E_{el}} $ and $\bar G = \bar {\delta
 G}$. 
 
 Representing vector $\bf G$ as a sum of components "parallel"
 and "normal" to the vector $\bf E$: ${\bf G} = {\bf G}_E +
 {\bf G}_{\perp}$ one has $({\bf G}_E, {\bf E}) = ({\bf G}, {\bf E})$,
 $({\bf G_{\perp}}, {\bf E}) = 0$, and, finally,
 $$ {\bf G}_E = {({\bf G},{\bf E}) \over {\bar E}^2}{\bf E}
 = C {\bar {\delta G} \over \bar E}{\bf E}$$

 Decomposing in the same way the random vector $\vec \xi_i$ on the
components "parallel" and "normal" to $\bf E$ ($\xi_E$ and
$\xi_{\perp}$) and taking into account that  
$$ <\delta G>_E =
<({\bf G}_E + {\bf G}_{\perp},{\vec \xi}_E + {\vec \xi}_{\perp})>_E, $$ 
$$({\bf G}_E, {\vec \xi}_{\perp}) = ({\bf G}_{\perp}, {\vec \xi}_E) =
0$$ 
and $<({\bf G}_{\perp}, {\vec \xi}_{\perp})> = 0$ we finally have
$$<\delta G>_E = <({\bf G}_E, {\vec \xi}_E)> = C{\bar {\delta G} \over \bar E}
({\bf E},{\vec \xi }) = C {{\bar \delta G} \over {\bar E}}E $$
which corresponds to Eq.\ref{<G_i>}. It means that $\delta G$ has a
linear regression with respect to $E_{el}$.


\begin{references}
 \bibitem{Rogers84}
 C.T.Rogers and R.A.Buhrman, Phys. Rev. Lett. {\bf 53}, 1272 (1984).
 \bibitem{Ralls88}
 K.S.Ralls and R.A.Buhrman, Phys. Rev. Lett. {\bf 60}, 2434 (1988).
 \bibitem{Kogan84}
 A.M.Kogan and K.E.Nagaev, Solid State Commun. {\bf 49}, 387 (1984);
 A.Ludviksson, R.Kree, and A.Schmid, Phys. Rev. Lett.  {\bf 52}, 950
 (1984).
 \bibitem{K84a}
 V.I.Kozub, Soviet Phys. JETP {\bf 59}, 1303 (1984).
 \bibitem{K84b}
 V.I.Kozub, Sov.Phys.-Solid State {\bf 26}, 1851 (1984).
 \bibitem{K84c}
 V.I.Kozub, Soviet Phys. JETP {\bf 60}, 818 (1984).
 \bibitem{Galp89}
 Yu.M.Galperin, V.G.Karpov, and V.I.Kozub, Adv. in Phys. {\bf 38}, 669
 (1989).
 \bibitem{AHVP} W.A.Phillips, Rep. Progress in Physics {\bf 50}, 1657 (1987).
 \bibitem{Kondo}
 J.Kondo, Physica {\bf 84B}, 40 (1976); {\bf 124 B}, 25 (1984); {\bf
 126B}, 377 (1984); Yu.Kagan and N.V.Prokof'ev, Sov. Phys. - JETP {\bf
 63}, 1276 (1986).
 \bibitem{Dissreview}
 J.Kondo, in {\it Fermi surface effects}, edited by J.Kondo, Springer
 series in Solid State Sciences Vol. 77 (Springer-Verlag, Heidelberg,
 1988).
 \bibitem{Zimm} N.M. Zimmerman, B. Golding, and W.H. Haemmerle, Phys. Rev.
Lett. {\bf 67}, 1322 (1991). 
 \bibitem{nondiss} V.I. Kozub, A.M. Rudin, and H.R. Schober, Phys. Rev. B 
{\bf 52}, 12705 (1995).
 \bibitem{Alt89}
 B.L.Altshuler and B.Spivak, Sov. Phys. JETP Lett. {\bf 49}, 772 (1989).
 \bibitem{KR93}
 V.I.Kozub and A.M.Rudin, Phys. Rev. B {\bf 47}, 13737 (1993).
\bibitem{Alt95} B.L. Altshuler, N. Wingreen, Y. Meir, Phys. Rev. Lett. {\bf
75}, 769 (1995).  
\bibitem{Magfield}
 C.W.J. Beenakker and H. van Houten, in {\it Solid State Physics},
 edited by H. Ehrenreich and D. Turnbull (Academic, New York, 1991),
 Vol. 44, p. 1.
 \bibitem{Galp91}
 Yu.M.Galperin and V.I.Kozub, Sov. Phys. - JETP {\bf 73}, 179
 (1991).
 \bibitem{Hultman}
 K.L.Hultman, J.Holder and A.V.Granato, J. Phys.  (Paris) {\bf 42},
 753 (1981).
 \bibitem{K86}
 V.I.Kozub and I.O.Kulik, Sov. Phys. - JETP {\bf 64}, 1332
 (1986).
 \bibitem{Black81}
 J.L.Black, in: ``Glassy metals I'', ed. by H.J.G\"untherodt and
 H.Beck, Berlin, Springer, 1981.
 \bibitem{Yanson}
 I.K.Yanson, Sov. Phys. - Low Temp. {\bf 9}, 343 (1983).
 \bibitem{Zawadovskii} A. Zawadovski, Phys. Rev. Lett. {\bf 45}, 211 (1980);
K.Vladar and A. Zawadovski, Phys. Rev. B {\bf 28}, 1564 (1983); {\bf 28}, 1582
(1983); {\bf 28} 1596 (1983); G. Zarand and A.Zawadovski, Phys. Rev. Lett. {\bf
72}, 542 (1994). 
 \bibitem{Ralph}
 D.C.Ralph and R.A.Buhrman, Phys. Rev. Lett. {\bf 69}, 2118 (1992);
 D.C.Ralph, Ph.D. dissertation, Cornell University, 1993; D.C.Ralph,
 A.W.W.Ludwig, Jan von Delft, and R.A.Buhrman, Phys.  Rev. Lett. {\bf
 72}, 1064 (1994).
 \bibitem{kozub94} V.I.Kozub, J.Caro and P.A.M.Holweg, Phys. Rev. B,
 {\bf 50}, 15126 (1994)
 \end{references}
 \end{document}